\documentclass[twocolumn,showpacs,preprintnumbers,amsmath,amssymb,pra]{revtex4}

\usepackage{graphicx}
\usepackage{amsmath,amsthm,amssymb}
\usepackage{bm}
\usepackage{dcolumn}
\usepackage{amsmath,amsthm,amssymb}
\usepackage{array}
\usepackage{multirow}

\newcommand{\triplet}{a \,^3\Sigma_u^+}

\newcommand{\tripletex}{(1)\,^3\Sigma_g^+}
\newcommand{\vib}{v'}
\newcommand{\qme}{|\vib\rangle}
\newcommand{\qmf}{|i\rangle}
\newcommand{\clg}{1_{\textrm{g}}}
\newcommand{\cog}{0_{\textrm{g}}^-}

\begin{document}

\title{Hyperfine, rotational and Zeeman structure of the lowest vibrational
levels of the $^{87}$Rb$_2$ $\tripletex$ state}

\author{T. Takekoshi $^{1}$}
\author{C. Strauss $^{1,2}$}
\author{F. Lang $^{1}$}
\author{J. Hecker Denschlag  $^{1,2}$}
\affiliation{$^{1}$Institut f\"ur Experimentalphysik, Universit\"at Innsbruck, 6020 Innsbruck, Austria \\
$^{2}$ Institut f\"ur Quantenmaterie, Universit\"at Ulm, 89069
Ulm, Germany}
\author{Marius Lysebo $^{4,5}$}
\author{Leif Veseth $^{5}$}
\affiliation{$^{4}$Faculty of Engineering, Oslo University College, 0130 Oslo, Norway\\
$^{5}$ Department of Physics, University of Oslo, 0316 Oslo, Norway}

\date{\today}

\begin{abstract}
We present the results of an experimental and theoretical  study
of the electronically excited $\tripletex$ state of $^{87}$Rb$_2$
molecules.  The vibrational energies are measured for deeply bound
states from the bottom up to $v'=15$ using laser spectroscopy of
ultracold Rb$_2$ Feshbach molecules. The spectrum of each
vibrational state is dominated by a 47\,GHz splitting into a
$\cog$ and $\clg$ component caused mainly by a strong second order
spin-orbit interaction. Our spectroscopy fully resolves the
rotational, hyperfine, and Zeeman structure of the spectrum.  We
are able to describe to first order this structure using a
simplified effective Hamiltonian.

\end{abstract}

\pacs{37.10.Mn, 42.62.FI, 33.20.-t}

\maketitle

\section{Introduction} \label{sec:int}
Progress in the field of ultracold atomic and molecular  gases has
always been strongly linked to developments in molecular
spectroscopy. Photoassociation spectroscopy, for example, has been
important for studies of ultracold atomic collisions and for
production of ultracold molecules
\cite{Wei99,Jon06,Koh06,Sage2005}. In 2008, after carrying out
spectroscopic searches, several groups managed to produce cold and
dense samples of deeply bound molecules in well-defined quantum
states \cite{La08, Ni08, Dan08, Vit08, Dei08,Ospelkaus2010}. For
this, a variety of clever optical transfer and filtering schemes
were developed which involved electronically excited molecular
levels. These levels had to be properly chosen for high efficiency
and selectivity of molecule production. This will also be the case
for future experiments involving cold collisions
\cite{Ospelkaus2010b,Ni2010}, ultracold chemistry \cite{Sta06,
Kre05, Kre08}, and testing fundamental laws via precision
spectroscopy \cite{Ze08, DeM08, Ch09}. A detailed understanding of
the excited molecular potentials is therefore necessary. Very
recent work \cite{Bai11} investigated the spin-orbit-coupled $A
^1\Sigma_u^+$ and $b ^3\Pi_u$ states of Cs$_2$. In other work
 a detailed analysis of  {\em weakly bound} Rb$_2$ levels of the
excited  $1_g$ state close to the $5S_{1/2}+5P_{1/2}$ dissociation
limit is currently under way \cite{Ber11} (see also related work
in Fig. 13 of \cite{Jon06}).

Here we present measurements and analysis for {\em deeply bound}
($v' = 0 ... 15$) levels of the $\tripletex$ (5$P_{1/2}$ +
5$S_{1/2})$ state of $^{87}$Rb$_2$. This state is relevant for the
production of deeply bound molecules in the $a ^3\Sigma_u^+$ state
via stimulated Raman adiabatic passage \cite{La08, Win07}. The $a
^3\Sigma_u^+$ state is the energetically lowest triplet potential,
giving rise to long lived molecules which are of interest for cold
collision experiments. The levels of the $a ^3\Sigma_u^+$ state
have been mapped out and identified in detail in a recent
publication \cite{Str10}.
\begin{figure}[htbp]
\centering
\includegraphics[width = 0.47\textwidth]{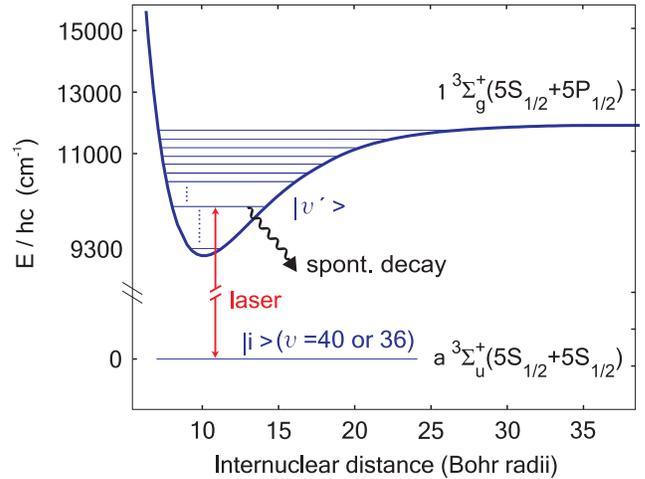}
\caption{Spectroscopy scheme. A tunable laser couples the Feshbach
molecule level $\qmf$ to an excited level $\qme$ with energy $E$
in the $\tripletex$ state. The excited state quickly decays
spontaneously. Level positions in the $\tripletex$ potential are
detected through resonantly enhanced loss of Feshbach molecules.
$h$ is Planck's constant and c is the speed of light.}
\label{fig:potential}
\end{figure}

The $\tripletex$ state is not easily accessible in conventional
setups since the  molecules in a  Rb$_2$ gas at ambient
temperatures
 are found in
the $X ^1\Sigma_g^+$ ground state. Electric dipole transitions
from this state to the $\tripletex$ state are forbidden by the
$g\leftrightarrow{}u$ symmetry selection rule. As a consequence it
has only quite recently become possible to explore in detail the
Rb$_2$ $\tripletex$ state.
 Lozeille {\it et al.}
\cite{L06} performed photoionization spectroscopy of ultracold
Rb$_2$ molecules produced by photoassociation in a magneto-optical
trap to resolve the large $\cog-\clg$ splitting of the vibrational
levels.
 Mudrich {\it et al.} \cite{Mu10} used pump-probe photoionization spectroscopy of
Rb$_2$ formed on helium nanodroplets to measure the vibrational
progression of deeply bound levels. Our work goes beyond this as
we fully resolve the rotational, hyperfine, and Zeeman structure
of the Rb$_2$ levels in the $\tripletex$ state.

The starting point for our experiments is an ultracold ensemble of
weakly bound Rb$_2$ Feshbach molecules in a well defined quantum
level, which has contributions from both  the $\tripletex$ and $X
^1\Sigma_g^+$ states. A scanning laser with sub-MHz short-term
linewidth drives a one-photon transition to levels in the
$\tripletex$ state (see Fig. 1). We obtain loss spectra for
various magnetic fields from 0\ to 1000\,G. The data are well
described by an effective Hamiltonian which contains terms for
molecular rotation as well as spin-spin, hyperfine, and Zeeman
interactions.

The article is organized as follows. Section \ref{sec:exp} of our
article presents the experimental setup  for the Rb$_2$
spectroscopy. In section \ref{sec:viblevel} we show typical
measured spectra and discuss their main features. Section
\ref{sec:ham} presents the model that we use to describe our data.
Section \ref{sec:res} explains the detailed level structure of our
data based on the model. We conclude with a summary and a short
outlook in section \ref{sec:con}.

\section{Experimental setup} \label{sec:exp}
\begin{figure}[tbp]
\centering
\includegraphics{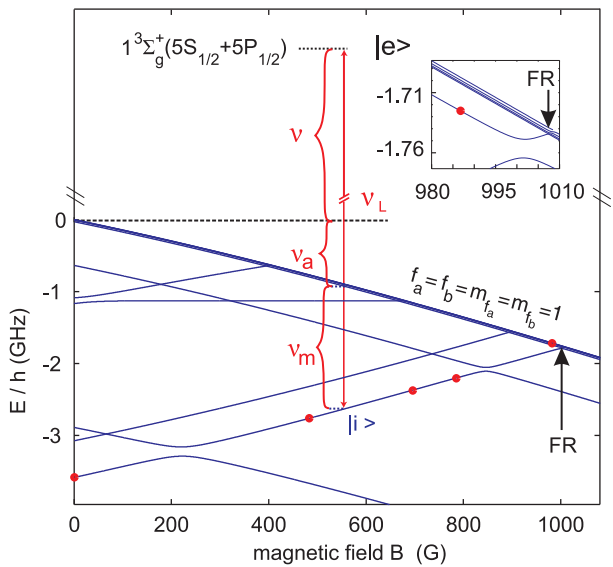}
\caption{Energy diagram of the Zeeman structure for some weakly
bound Rb$_2$ molecular states below the $|f_a=1, m_{f_a} =
1\rangle$ + $|f_a=b, m_{f_b} = 1\rangle$ atomic asymptote.  Dots
mark the levels and magnetic fields used for the spectroscopy. For
simplicity, throughout the paper we report excitation frequencies
$\nu$ with respect to the two-atom threshold at 0\,G. These are
obtained by subtracting the known atomic Zeeman shift $\nu_a$ and
the known molecular binding frequency $\nu_m$ from the excitation
frequency $\nu_L$. The inset zooms into the region near the
Feshbach resonance at 1007.4 G, marked with an arrow and ''FR".}
\label{fig:feshbach}
\end{figure}

 Feshbach molecules in state $\qmf$ are
irradiated by a pulse of light from a continuous-wave laser the
wavelength of which is slowly scanned over the range 1000 to 1050
nm (Fig.\,\ref{fig:potential}). The laser beam has an $1/e^2$
intensity waist radius of 130\,$\mu$m at the molecular sample.
It is linearly polarized
along the magnetic bias field $\mathbf B$ 
(parallel to gravity) and thus can only induce $\pi$ transitions.
The light pulse typically lasts for 50\,ms and has a rectangular
temporal shape. When resonant with an $\qmf-\qme$ transition, our
laser induces losses in the Feshbach population due to excitation
to $\qme$ and subsequent fast (ns) decay to unobserved states. For
each data point, a new ensemble of Feshbach molecules has to be
prepared, a process which typically requires 28\,s.

The preparation of the Feshbach molecules is described in detail
in \cite{Tha06, La08a}.  In brief, following laser cooling and
evaporative cooling, we trap an ultracold atomic cloud of 3$\times
10^{5}$ $^{87}$Rb atoms close to quantum degeneracy in a 3D
optical lattice. A sizeable fraction ($\approx 1/3$) of the
lattice sites are doubly occupied and the atoms are trapped in the
lowest Bloch band. Afterwards, we ramp over a Feshbach resonance
at a magnetic field of 1007.4\,G (1\,G = 10$^{-4}$\,T) (see
Fig.\,\ref{fig:feshbach}) \cite{Vol03} to obtain $3\times10^4$
Feshbach molecules with no more than a single molecule per lattice
site. The lattice depth for the Feshbach molecules is deep enough
to prevent the molecules from colliding with each other, and we
observe lifetimes of up to a few hundred ms.

In order to investigate the Zeeman structure of the levels in the
$1 ^3\Sigma_g^+$ state, we carry out the spectroscopy at various
magnetic fields between 0 and 1000\,G. Thus, after we have
produced the Feshbach molecules, the magnetic field is ramped down
from  1007.4\,G to its chosen value. When doing so, one has to
keep track of the quantum state $\qmf$ of the Feshbach molecules
as well as their binding energy as they both change in general
with magnetic field. Fig.\,\ref{fig:feshbach} shows a number of
relevant weakly bound molecular energy levels which exhibit many
avoided crossings. When ramping over such a crossing, the
molecules could potentially end up in two different quantum
levels. This would be problematic for the spectroscopy and is
avoided as described below.

A large fraction of our measurements are carried out at 986.8\,G
with a binding energy of 22.7\,MHz$\times h$ (see inset of
Fig.\,\ref{fig:feshbach}).  The corresponding Feshbach state is
best described in the atomic basis and correlates to
$|v=40,(f_{a}=1, f_{b}=1) f = 2, l=0, F=2, M=2\rangle$ at low
magnetic fields. Here $v$ is the vibrational quantum number of the
$a ^3\Sigma_u^+$ state, and $f_a$,$f_b$ are the total angular
momenta of atoms $a$ and $b$. The sum of these angular momenta,
$\mathbf{f}=\mathbf{f_a}+\mathbf{f_b}$, and the rotational angular
momentum of the atoms $\mathbf{l}$ couple to form the total
angular momentum $\mathbf{F}$. At high magnetic field, $F$ is no
longer a good quantum number; only its projection $M$ onto the
$z-$axis remains good. As an estimate of this effect, we note, for
example, that at 986.8\,G the expectation values \cite{expectval}
for $f_a$ and $f_b$ become about 1.5.

For magnetic fields below 986.8\,G we use a molecular level  which
correlates to $|v=36,(f_{a}=2, f_{b}=2) f = 2, l=0, F=2,
M=2\rangle$ at low magnetic fields. This is the diagonal line in
Fig. 2 going from the point (
$B = 0$, $E / h \approx -3.6 $ GHz) to the Feshbach resonance at
threshold. It exhibits several avoided crossings with other
molecular levels.  When ramping down the magnetic field, these
avoided crossings are crossed using a adiabatic radiofrequency
transfer method \cite{La08a}. In order to count the molecules
remaining after the spectroscopy pulse, we retrace our path back
to the Feshbach resonance at 1007.4 G where the molecules are
dissociated via a reverse Feshbach magnetic field sweep
\cite{Tha06} and imaged as atoms using standard absorption imaging
techniques \cite{Kett99}.

The spectroscopy laser is either a Ti:sapphire or a
grating-stabilized diode laser, both of which can be locked to an
optical cavity using the Pound-Drever-Hall scheme. The optical
cavity is stabilized with respect to an atomic $^{87}$Rb line.
While the unlocked lasers typically drift a few MHz within one
experimental cycle, the cavity lock leads to a stability better
than 1 MHz. Locking was only necessary for resolving a few weak
lines.  The laser frequency was determined using a commercial
wavemeter (HighFinesse WS7) within seconds after the laser pulse.
The wavemeter has an accuracy of 60\,MHz after calibration, which
is done daily using an $^{87}$Rb line. Over the length of only a
few experimental cycles (5\,minutes) it typically drifts less than
10\,MHz which sets a precision limit on relative line positions
when working with an unlocked laser.

We choose the $|f_a=1, m_{f_a} = 1\rangle$ + $|f_a=b, m_{f_b} =
1\rangle$ dissociation threshold at 0\,G as the energy reference
($E=0$).  We thus substract from the measured laser excitation
energy $h\nu_L$, both the bound state energy of the Feshbach level
$h \nu_m$ as well as the Zeeman energy $h \nu_a$ of the free atoms
(Fig.\,\ref{fig:feshbach}), which are both well known
\cite{Str10}.

\section{Experimental observations}
\begin{figure}[tbp]
\centering \includegraphics{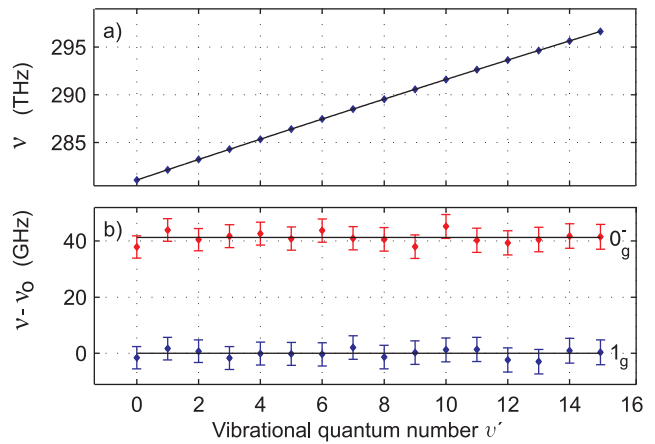} \caption{Vibrational
ladder for the $\tripletex$ state. a) Excitation  frequencies
$\nu$ of the vibrational states with quantum number $\vib$. The
solid line is a quadratic fit to the data. b) Zoom into the
vibrational levels which are split into a $\cog$ and a $\clg$
component. The spectra are measured at low resolution and high
power.  A frequency offset $\nu_o(\vib)$ is subtracted from the
excitation frequency $\nu$ for each data point. The offset
$\nu_o(\vib)$ is a  quadratic fit  to the $\clg$ data. Note the
different energy scales of a) and b).} \label{fig:vibladder}
\end{figure} \label{sec:viblevel}

In a first set of experiments, we have mapped out the vibrational
ladder of the $\tripletex$ state from $\vib=0$ to $\vib=15$ at low
resolution (Fig.\,\ref{fig:vibladder}a). We used the Ti:sapphire
laser with an available power of a few hundred mW such that we
only observed broad lines with a typical width of several GHz. The
magnetic field was set to 986.8\,G. The vibrational ground state
of the $\tripletex$ state has an excitation frequency of
281.1\,THz with respect to $\qmf$, corresponding to a laser
wavelength of 1067nm. The vibrational splitting between the two
lowest levels, $\vib=0$ and $\vib=1$, is about 1.2\,THz. The solid
line in Fig.\,\ref{fig:vibladder}a is a quadratic fit in $v$
allowing for a slight anharmonicity in the vibrational ladder. The
data lead to the same Morse potential as given by Mudrich et al.
\cite{Mu10}. We find that each vibrational level is a doublet with
a splitting of about 47\,GHz \cite{lambdapar} (see
Fig.\,\ref{fig:vibladder}\,b). The large error bars of several GHz
reflect the crudeness of this first measurement where we do not
resolve the substructure of each of the doublet components.

\subsection{Splitting of the vibrational levels into $\cog$ and $\clg$ components}
The 47\,GHz splitting of the vibrational levels clearly cannot  be
explained by the rotational, hyperfine, or Zeeman interactions
because they would be too small. Estimating the molecular
hyperfine and Zeeman energies from those for $^{87}$Rb atoms, we
expect such contributions to be at most 14 GHz. It turns out that
the large splitting comes from a strong effective spin-spin
coupling of the electrons. Although there is direct spin-spin
coupling, the main contribution is second order spin-orbit
coupling, which is resonantly enhanced by the nearby $(1)^1\Pi_g$
state \cite{L06}. Experimentally, these two contributions cannot
be separated. In its microscopic form, the effective spin-spin
interaction Hamiltonian reads
\begin{align}
H_\text{ss} = -C\frac{3 (\mathbf{s}_1\cdot \mathbf{r}_{12})
(\mathbf{s_2}\cdot \mathbf{r}_{12}) - (\mathbf{s}_1\cdot
\mathbf{s}_2)\mathbf{r}_{12}^2}{r_{12}^5},
\end{align}
where $C$ is a constant, $\mathbf{s}_{1,2}$ are the electron spin
operators, and $\mathbf{r}_{12}$ is the relative position vector
of the two electrons. It can be shown \cite{Kramers1,Kramers2}
that for $\Sigma$ states, $H_\text{ss}$ can be simplified to
\begin{align}
H_\text{ss} = 2 \lambda [(\mathbf{n}\cdot \mathbf{S})^2 -
\mathbf{S}^2/3].
\end{align}
Here,  $\lambda$ is effective molecular parameter for the
spin-spin interaction which has to be determined for the studied
vibrational level. $\mathbf{n}$ is the unit vector along the
internuclear axis, and $\mathbf{S}$ is the total electronic spin
operator with $\mathbf{S} = \mathbf{s}_{1} + \mathbf{s}_{2}$.
Since the term $\mathbf{S}^2 / 3$ merely results in an overall
offset, it will be ignored. Thus the spin-spin interaction takes
the form
\begin{align}
H_\text{ss} = 2 \lambda(\mathbf{n}\cdot \mathbf{S})^2.
\label{eq:ss}
\end{align}
A strong $H_\text{ss}$ couples the electronic spin to the
internuclear axis, making its projection $\Sigma = 0, 1$ a good
quantum number. Thus, for our $\tripletex$ state $\Omega = \Lambda
+ \Sigma = \Sigma$ is also a good quantum number. Here $\Lambda$
and $\Omega$ are the projections of the total  electronic orbital
angular momentum $\mathbf{L}$ and the total angular momentum
$\mathbf{J} = \mathbf{L} + \mathbf{R}  + \mathbf{S}$ on the
internuclear axis, where $\mathbf{R}$ is the rotational angular
momentum of the nuclei. The eigenvalues of $H_\text{ss}$ are
$2\lambda\Sigma^2 $. This means that the splitting between the
$\Omega = 0$ and $\Omega = 1$ states is 2$\lambda$.
As will become clear later, the more deeply bound component of the
observed doublet structure has $\clg$ character and the other one
has $\cog$ character (Fig.\,\ref{fig:vibladder}\,b), where we use
Hund's case (c) notation $|\Omega|_{g/u}$.

\subsection{Spectra of the $\cog$ and $\clg$ states}
\label{sec:0g1gOverview} 
\begin{figure}[tbp]\centering
\includegraphics{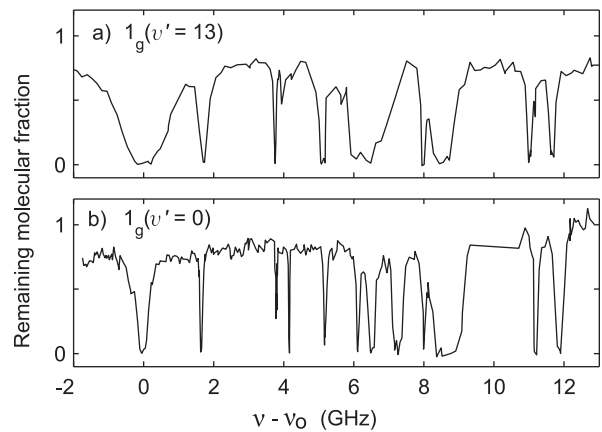}
\caption{The $\clg$ spectra of two different  vibrational levels
at 986.8\,G. For each spectrum, we subtract a frequency offset
$\nu_o$ from the excitation frequency $\nu$ such that the most
deeply bound line is situated at $0$\,GHz. a) $\vib=13$, $\nu_o =
294626.4$\,GHz.  b) $\vib=0$, $\nu_o = 281068.2$\,GHz. In these
two scans, not all lines that we eventually observed are
resolved.} \label{fig:v0v13}
\end{figure}

\begin{figure}[tbp]
\centering \includegraphics{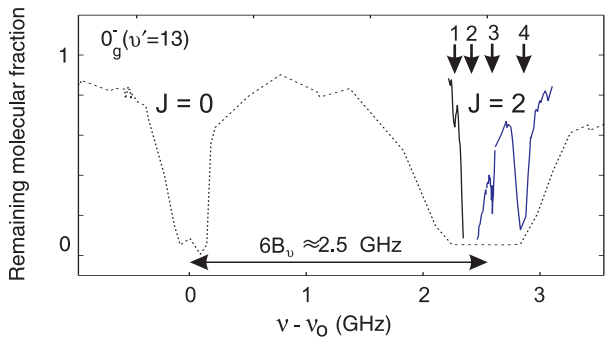} \caption{Scans of $\cog
(\vib=13)$ lines  in the $\tripletex$ state at a magnetic field of
986.8\,G.  The dotted line corresponds to a laser power of 0.1\,mW
and a pulse duration of 200\,ms while the solid line corresponds
to a laser power of 0.1\,mW and a 1\,ms pulse duration.   The
frequency offset is $\nu_o =294671.0$\,GHz. The line at
$\nu-\nu_o$ = 0\,GHz corresponds to $J = 0$, while the 4 lines
located around 2.5\,GHz have $J = 2$. Due to saturation these 4
lines are not resolved for the 200\,ms pulse. } \label{fig:v13_0g}
\end{figure}

The $\cog$ and $\clg$ states have a rich
substructure  which we are able to resolve by lowering the power
of the laser to about 0.1\,mW.  Figures \ref{fig:v0v13} and
\ref{fig:v13_0g} show loss spectra for $\clg$ ($\vib=0,\,13$) and
$\cog$ ($\vib=13$), respectively. While the $\clg$ manifold is
spread out over 12\,GHz, the $\cog$ manifold is narrower (3\,GHz)
and has fewer lines. The 12 observed lines of the $\clg$ manifold
in Fig. \,\ref{fig:v0v13} show no obvious pattern.  The structure
is the result of an interplay of rotational, hyperfine, and Zeeman
interactions.  It is one of the main goals of this work to
understand this structure and to identify the individual lines.

A $\clg$ spectrum typically consists of roughly 300 points where
each point  corresponds to one production and measurement cycle.
The lines that we observe vary markedly in width. This indicates a
strong variation of the laser-induced coupling between $\qme$ and
$\qmf$. Spectra for different vibrational levels are similar
(Fig.\,\ref{fig:v0v13}).

Compared to the typical step size of 40 MHz in
Fig.\,\ref{fig:v0v13},  the 12 MHz natural linewidth  of the
molecular levels (given by twice the atomic linewidth) is
relatively small. Thus, it is possible that some weak lines are
not always detected, especially when they are located on the
shoulder of a power-broadened line. We thus carried out a number
of scans with various step sizes, laser powers, and pulse times,
testing for consistency and checking theoretical predictions.  For
the $\clg$ spectra we found 18 lines in total (Section
\ref{sec:res}).

Our $\cog$ spectrum is considerably simpler than the $\clg$
spectrum.  There are 5 lines arranged in a doublet-like structure
with a splitting of $\approx2.5$\,GHz (Fig.\,\ref{fig:v13_0g}).
This splitting is due to rotation of the molecule \footnote{Here
we neglect the centrifugal distortion since the centrifugal
distortion constant $D$ is a factor 10$^6$ smaller than
$B_{\vib}$, and we only work at low rotations.},

\begin{align}
H_{\text{rot}} =B_{v'}\mathbf{R}^2 / \hbar^2,\label{eq:rotJ}
\end{align}

where $B_{v'}$ is the rotational constant and $\hbar$ is the
reduced Planck constant. The hyperfine and Zeeman contributions
are much weaker here. As we will show in section \ref{sec:ham}
this can be explained due to the vanishing of the projections of
total spin ${\bf S}$ and total orbital angular momentum ${\bf L}$
of the electrons on the internuclear axis. The rotational constant
$B_{v'}$ is
\begin{align}
B_{v'} = \frac{\hbar^2}{2\mu} \biggl \langle v'\biggl |
\frac{1}{r^2} \biggl |  v'\biggl \rangle,
\end{align}

where $|v'\rangle$ is the ket for the vibrational wave function,
$r$ is the nuclear separation, and  $\mu$ is the reduced mass.
From previous investigations, \cite{Par01,L06} we expect
$B_{v'=13}/ h =412$\,MHz.  Due to the weakness of the hyperfine
and Zeeman interactions, the angular momentum ${\bf J} $
 is conserved.
Apart from an offset the rotational energy is then determined from
Eq. (\ref{eq:rotJ}) to be \cite{Lefebvre2004}
\begin{align}
E_{\text{rot}} =B_{v'} \ J (J +1).
\end{align}

We observe the lines $J=0$ and $J=2$ separated by $6B_{v'} / h
\approx2.5$\,GHz (see Fig. 5).  The rotational level $J=1$ is not
accessible because total parity (under inversion of all electron
and nuclear coordinates) has to change in the optical transition.
The parity for the Feshbach molecule is $(-1)^l = +1$ while the
parity of the states in the $0_g^-$ manifold of the $\tripletex$
state is $(-1)^{J+1}$.  The $J=2$ feature has a substructure of
four lines which is due to residual hyperfine and Zeeman
interaction. This substructure will be discussed in detail in
section \ref{sec:res} hand in hand with the analysis of our
theoretical model.

\section{Effective Hamiltonian and evaluation of its parameters}
\label{sec:ham} In the following we present the diatomic model
Hamiltonian which  we use to describe the observed energy levels
within the $\vib = 13$ vibrational manifold.  Spectra in other low
lying vibrational manifolds are similar (Fig. 4), and can be
described by the same Hamiltonian with slightly adjusted
parameters.  We can thus ignore electronic terms and vibrational
motion.  The Hamiltonian can then be written,

\begin{align}
H =   H_{\text{ss}}+ H_{\text{rot}} + H_{\text{hf}} +
H_{\text{Z}}+ H_{\text{sr}}. \label{eq:hamilton}
\end{align}

$H_{\text{ss}} = 2\lambda(\mathbf{n}\cdot \mathbf{S})^2$ is the
effective spin-spin operator given in Eq. (3) which, as previously
discussed, leads to the large splitting into the $\cog$ and $\clg$
components. $H_{\text{rot}} = B_{v'}\mathbf{R}^2 / \hbar^2$ is the
Hamiltonian for molecular rotation from Eq. (4).  The terms for
hyperfine interaction $H_{\text{hf}}$, Zeeman interaction
$H_{\text{Z}}$, and finally spin-rotation interaction
H$_{\text{sr}}$ are described in the following. Since the goal in
this article is to get a first understanding of the experimentally
observed spectra, we will in general simplify the interaction and
only take into account terms of leading order.

\subsection{The hyperfine interaction}
A general ansatz for the hyperfine Hamiltonian in  Hund's case a)
and b) is of the form \cite{Tow55}

\begin{align}
H_{\text{hf}}= a \ \Lambda \ \mathbf{I} \cdot \mathbf{n}  + (b_F-
{1 \over 3} c) \ \mathbf{I} \cdot\mathbf{S}+c \
(\mathbf{I}\cdot\mathbf{n})\ (\mathbf{S}\cdot\mathbf{n}).
\label{eq:defa}
\end{align}

Here, $\mathbf{I}$ is the operator for the total nuclear spin. The
first term describes the interaction of the electronic orbital
angular momentum with the nuclear spin. However, since $\Lambda =
0$, this term will not contribute.  The Fermi contact parameter is
$b_F$, while $c$ is called the anisotropic hyperfine parameter.
For $\Sigma$-states, we expect $c\ll b_F$ (\cite{Tow55}, page
196). These two parameters could in principle be calculated
ab-initio, but we have used them as free fit parameters. We note
that the total nuclear spin I is a good quantum number for the
given Hamiltonian.

\subsection{The Zeeman interaction}
Because we carry out measurements at high magnetic fields, the
Zeeman interaction plays an important role.  The main contribution
to the Zeeman interaction comes from the electrons while
contributions from the nuclear spins and molecular rotation (as
well as second order effects treated in \cite{Veseth}) are much
smaller and are neglected here. Furthermore, since $\Lambda$
vanishes in the $\tripletex$ state, the Zeeman interaction due to
the total orbital angular momentum of the electrons, $\mu_B
\mathbf{L} \cdot \mathbf{B}/ \hbar$
is also negligible to first order. (Here, $\mu_B$ is the Bohr
magneton). The only remaining term is

\begin{align}
H_Z = \mu_B g_S  \mathbf{S} \cdot \mathbf{B}/\hbar ,
\label{eq:zeeman}
\end{align}

where $g_S$ is the electron g-factor.  We note that there is no
free fitting parameter in the Zeeman Hamiltonian for adjusting the
model to the measured data. While the contribution from the Zeeman
interaction will in general be large for the $\clg$ lines, it will
be small for the $\cog$ lines because here the spin projection
onto the internuclear axis $\Sigma = 0$.

\subsection{The spin-rotation interaction}
In principle, we could also include a spin-rotation interaction through the effective Hamiltonian

\begin{align}
H_{\text{sr}} = \gamma_{v'} \mathbf{N}\cdot \mathbf{S}.
\label{eq:sr}
\end{align}

where $\mathbf{N} = \mathbf{L}+\mathbf{R}$ and $\gamma_{v'}$ is
the spin-rotation coupling constant. However, since $\gamma_{v'}$
is typically a small fraction of the rotational constant $B_{v'}$
\cite{Lefebvre2004}, the spin-rotation interaction only represents
an insignificant correction to the energy levels in the present
system and is thus neglected.

\subsection{Fit procedure and evaluation of molecular parameters}
According to our model Hamiltonian in Eq. (\ref{eq:hamilton})
there are  four adjustable parameters: the rotational constant
$B_{\vib}$, the spin-spin splitting parameter $\lambda$, the
Fermi-contact parameter $b_F$, and the anisotropic hyperfine
parameter $c$. As discussed before in section
\ref{sec:0g1gOverview}, the rotational constant $B_{\vib}$ should
be close to 412\,MHz based on previous work, and in agreement with
our analysis of the $\cog$ spectrum in Fig. 5.  The spin-spin
parameter $\lambda$ is  determined by the splitting of the $\clg$
and $\cog$ manifolds to be $2\lambda = 47$\,GHz. This leaves only
$b_F$ and $c$ as completely free parameters.  We determined all
parameters from fits of the model to the experimental data using a
nonlinear Levenberg-Marquardt method. For the calculations, the
Hamiltonian in Eq. (\ref{eq:hamilton}) is expressed  in terms of
matrix elements in a Hund's case (a$_{\alpha}$) basis, where the
basis states are of the form

\begin{align}
|\Lambda, S, \Sigma, I_1, I_2, \Omega_{I_1},\Omega_{I_2}, F,
\Omega_F, M\rangle, \label{eq:basis}
\end{align}

where $\Omega_F = \Lambda + \Sigma + \Omega_{I_1} + \Omega_{I_2}$
is  the projection of the total angular momentum $F$ on the
internuclear axis. $I_1$ =  $I_2$ = 3/2 are the nuclear spins of
the two nuclei and $\Omega_{I_1}$, $\Omega_{I_2}$ are their
projections onto the internuclear axis. We note that the basis set
(\ref{eq:basis}) is chosen larger than necessary for our purpose.
It can be conveniently used to investigate also Hamiltonians where
the total nuclear spin $I$ ($\mathbf{I} = \mathbf{I_1} +
\mathbf{I_2}$) might not be a good quantum number. For the
analytical expressions of the matrix elements as a function of the
quantum numbers of Eq. (\ref{eq:basis}), we refer the reader to
\cite{Marius09}. The Hamiltonian matrix is then numerically
diagonalized to obtain the eigenvalues and eigenstates. Included
in this calculation are all hyperfine states in the $^3\Sigma_g^+
(v'=13)$ electronic state with total angular momentum of up to $F
= 10$. Experimentally, we have only observed states with $F<7$.
The parameters $b_F $ and $c$ are determined in terms of
combination $b_F + {2 \over 3} c$ and $b_F -  {1 \over 3} c$,
which correspond to contributions diagonal and off-diagonal in
$\Sigma$ and $\Omega_I= \Omega_{I_1} + \Omega_{I_2}$,
respectively. The diagonal term can be directly read off from Eq.
(\ref{eq:defa}).

\begin{align}
\begin{split}
H_{\text{hf}}^{diag}&= (b_F- {1 \over 3} c)\, \Omega_I \Sigma +
c\, \Omega_I  \Sigma \\&= (b_F+ {2 \over 3} c)\,  \Omega_I
\Sigma.
\end{split}
\end{align}

Our final analysis gives $B_{\vib} = 412$\,MHz, $2\lambda =
47$\,GHz and $b_F + {2 \over 3} c = 832$\,MHz. We find, however,
that $b_F - {1 \over 3} c$ is not precisely determined in our
analysis. Experimentally, the parameter $b_F - {1 \over 3} c$
could, in principle, be best determined from the $\cog$ spectrum
because the contribution to the hyperfine interaction is purely
off-diagonal with respect to $\Sigma $. However, the energy levels
depend only weakly on this parameter, and our measurements
indicate a value between $(200-1000)$\,MHz. Higher precision than
that reached in our measurements is required in order to better
estimate $(b_F-\frac{1}{3}c)$.

\section{Assignment of lines of the $\cog$ and $\clg$
spectra }\label{sec:res} With the help of our theoretical model,
we can now identify the  individual lines of the  $\cog$ and
$\clg$ spectra and understand the physical origin of the
substructure of these spectra. Since the Feshbach molecules have
total magnetic quantum number $M = 2$, and since we use an optical
$\pi$ transition, we only observe excited levels with $M = 2$ and
thus $F\geq2$. From our discussion so far, we have the following
good quantum numbers for our levels in the $\tripletex$ state: $M
= 2, \Lambda = 0; \Sigma = 0, 1; \Omega = 0, 1; S = 1.$
Furthermore the total nuclear spin $I$ is a good quantum number in
our Hamiltonian and as we will see in the next section can take
the values $I = 1, 3$. At low magnetic field, $F$ is also a good
quantum number. With these and the additional quantum number $J$,
which (as we will show) is good for the $\cog$ states and becomes
good for high-rotational $\clg$ states, we will be able to
identify all lines.

\subsection{$\cog$ Zeeman spectrum}

\begin{figure}[tbp]
\includegraphics{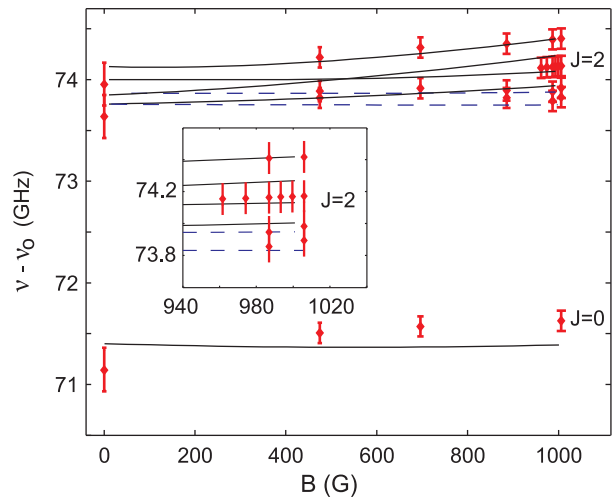} \caption{\label{fig:omega0} Zeeman diagram
of the $\cog$ ($v' = 13$) state. An offset $\nu_o = 294600$\,GHz
is subtracted  from the excitation frequency $\nu$. The lines are
calculations.
 Dashed lines indicate levels with $F > 3$ that cannot
 be excited in our experiment at 
$B =0$
 due to the selection rule on $\Delta F = 0,
\pm 1$. Table \ref{tbl:qn0} gives the quantum numbers at 0\,G and
1005.8\,G.  The inset is a zoom into the closely spaced lines
around 1000 G.}
\end{figure}

Figure \ref{fig:omega0} shows experimentally observed energy
levels for various magnetic field strengths along with the
calculations (dashed and solid lines).  The Zeeman effect is quite
weak as levels shift no more than a couple hundred MHz over a
range of 1000 G. This is consistent with our previous assertion in
section IIIb that the Zeeman interaction
($\propto \mathbf{S}\cdot \mathbf{B}$) is small because the
expectation value of the electron spin vanishes in any projection
direction for a state with $\Sigma = 0$. An identical argument
holds for the hyperfine interaction in Eq. (8). The observed
residual splitting of the lines in Fig. 6 is mainly due to second
order contributions of the Zeeman and hyperfine interactions.  The
$\cog$ spectrum is dominated by the rotational splitting
$B_{v'}J(J+1)$.  This overall trend is well described by the
theory (Fig. \ref{fig:omega0}).  A slight discrepancy can be
observed in the rotational splitting which seems to lie somewhat
outside the experimental error bars.  Note that at 0~G,
experimental data are less accurate than for the other magnetic
fields because of technical stability issues when we transfer the
Feshbach molecules from 1007\,G to low magnetic fields.
Uncontrolled magnetic field drifts lead to fluctuations in the
transfer efficiency across the avoided crossings from shot to
shot, which made the measurements noisy.

We now study in detail the quantum numbers of the observed levels.
Based on the exchange symmetry of the nuclei and other fundamental
symmetries of the $\tripletex$ state, one can show that the $J =
0$ and $J = 2$ levels have either total nuclear spin $I$ = 1 or 3.
$\mathbf{J}$ and $\mathbf{I}$ couple to form the total angular
momentum $\mathbf{F}$. For $J = 0$ we obtain $F = 1, 3$, of which
only $F = I = 3$ is observed (the single line at
$\approx\,$294671\,GHz in Fig. 6).  This is because we only detect
levels that have a total magnetic quantum number $M = 2$. Coupling
$J = 2$ with $I=1$ ($I=3$) results in states with $F=2,\,3$
($F=2,\,3,\,4,\,5$), where we have omitted $F = 1$ states, since
again they cannot be observed. These six levels are plotted as
lines in Fig. 6 and also listed in table \ref{tbl:qn0} together
with measured data. As can be read off from the expectation values
in the table,  for low magnetic fields the levels  in the $\cog$
component are indeed well described by the quantum numbers $
J,I,F. $ At 1000 G, $F $ is not good anymore.

In Fig. \ref{fig:omega0}, two of the calculated lines are dashed.
They correspond to $F = 4, 5$ which are not observable at 0G
because of the selection rule $\Delta F = 0,\pm 1$. For larger
magnetic fields, the Zeeman interaction mixes states with
different $F$ and the two levels that correlate with $F = 4, 5$ at
0 G become observable. As the magnetic field is increased, some
lines cross. Because $I$ is exactly a good quantum number for our
specific Hamiltonian $H$, crossings between levels with different
$I$ quantum numbers are not avoided.

From Fig. 6 as well as from Table I, we see that apparently the
two $I = 1$ lines are not observed.  Closed coupled-channel
calculations \cite{privTiemann, Str10} show that our Feshbach
molecules only have a few percent $I = 1$ character.  The
selection rule $\Delta I = 0$ thus makes it clear that $I = 1$
lines will be suppressed in the spectrum. We also know that the
Feshbach state is a superposition of electronic singlet and
triplet states.  For $l = 0$ one finds from symmetry arguments
(inversion symmetry, rotational symmetry, exchange symmetry,
reflection symmetry) that the triplet component has nuclear spin
$I = 1,3$ while the singlet component has $I = 0, 2$. Further, for
$l = 0$, the triplet component of the Feshbach state in the Hund's
case (c) basis has $J = 1$, $\Omega = 0$ while the singlet
component has $J = 0$. Only the electronic triplet part of the
Feshbach state will contribute to the optical transition to the
$\tripletex$ $\cog$ state, fulfilling the standard Hund's case (c)
$\Omega = 0 \leftrightarrow \Omega = 0$ selection rules $\Delta J=
\pm 1, \Delta S = 0,$ and $\Delta I = 0$.

\begin{figure*}[t]
\includegraphics{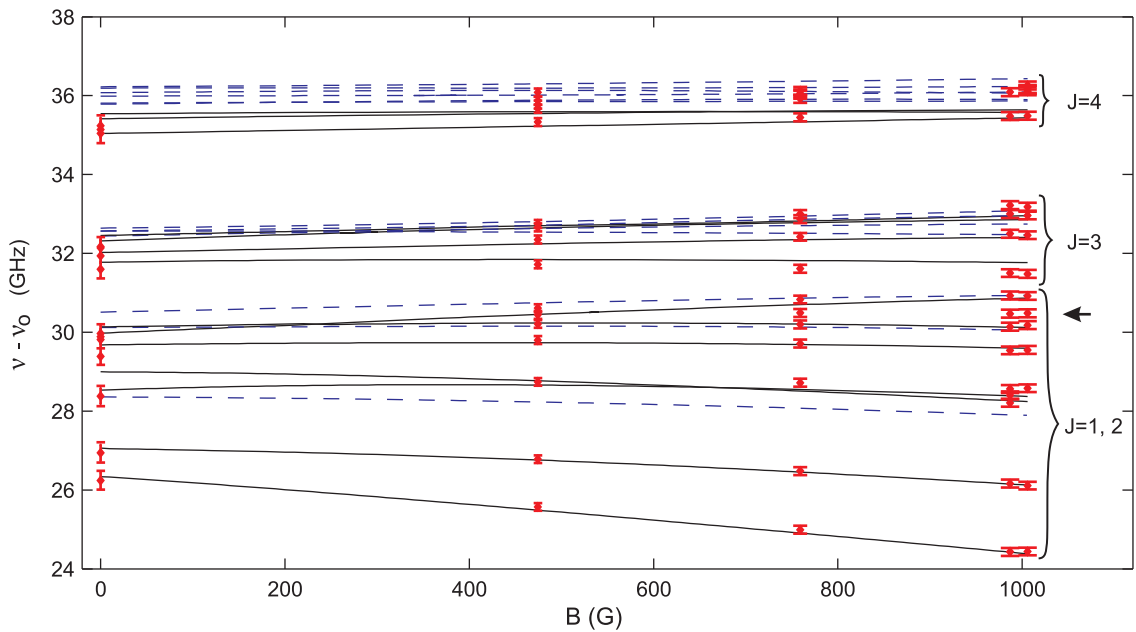}
\caption{Zeeman diagram of the $\clg$ ($v' = 13$) state.  An
offset $\nu_o = 294600\,\mathrm{GHz}$ is subtracted from the
excitation frequency $\nu$. The lines are calculations. Dashed
lines indicate levels with $F > 3$ that can not be observed in our
experiment at 
$B = 0$ due to the selection rule $F = 0, \pm 1$. The arrow points
to an experimentally observed line that is not well described by
the calculations. Table \ref{tbl:qn1} gives the corresponding
quantum numbers at 0\,G.}\label{fig:omega1}
\end{figure*}

\begin{table}[tbp]
\begin{center}
\renewcommand{\multirowsetup}{\centering} 
\begin{tabular}{|c||c|c|c|c||c|c|c|c|}
\hline
&  \multicolumn{4}{c||}{$B=0$
\,G} & \multicolumn{4}{|c|}{$B 
=1005.8$\,G}\\[1ex]
\hline
\multirow{2}{*}{$\langle J\rangle$} & \multirow{2}{*}{$\langle I\rangle$} & \multirow{2}{*}{$\langle F\rangle$} & $\nu_{T} - \nu_o$ & $\nu_{E} -\nu_o$ & \multirow{2}{*}{$\langle I\rangle$} & \multirow{2}{*}{$\langle F\rangle$} & $\nu_{T} - \nu_o$ & $\nu_{E} -\nu_o$\\
& & & (GHz) & (GHz) &  & & (GHz) & (GHz)\\[0.5ex]
\hline
0.0 & 3 & 3.0 & 71.44 & 71.14 &   3.0 &   3.0 & 71.42 & 71.63\\
2.0 & 3 & 5.0 & 73.84 & n.o.  &   3.0 &   4.2 & 73.83 & 73.83\\
2.0 & 1 & 3.0 & 73.84 & n.o.  &   3.0 &   3.7 & 73.95 & 73.92\\
2.0 & 1 & 2.0 & 73.92 & n.o.  &   1.0 &   2.9 & 74.00 & n.o.\\
2.0 & 3 & 4.0 & 73.94 & n.o.  &   3.0 &   3.3 & 74.13 & 74.13\\
2.0 & 3 & 3.0 & 74.06 & 73.64 &   1.0 &   2.1 & 74.27 & n.o.\\
2.0 & 3 & 2.0 & 74.17 & 73.95 &   3.0 &   2.8 & 74.42 & 74.40\\
\hline
\end{tabular}
\caption{\label{tbl:qn0} Calculated and measured $\cog$
$(\vib=13)$  excitation frequencies of the levels shown in Fig.
\ref{fig:omega0} at 0\,G and 1005.8\,G. The offset from the
excitation frequency $\nu$ is $\nu_o = 294600$\,GHz. The
subscripts $_T$ and $_E$ denote theoretical and experimental
values.  The expectation values for quantum numbers $J$, $I$ and
$F$ are also given. All levels have $S=1$, $\Sigma=0$, and $M=2$.
Levels with $I = 1$ are not observed (n.o.). The absolute accuracy
for the measured line
positions at $B = 0$ 
and $1005.8$\,G is about 200\,MHz
and 70\,MHz, respectively. The accuracy for the relative positions
of the lines, however, is in general higher, e.g. $\approx$20\,MHz
for the J = 2 lines at $B 
 = 1005.8$\,G.}
\end{center}
\end{table}

\subsection{$\clg$ Zeeman spectrum }
The experimental $\clg$ data at various magnetic fields together
with the calculated energies are shown in Fig.\,\ref{fig:omega1}.
There is good agreement of the overall structure of the calculated
levels and the observed data. Our calculations (Table
\ref{tbl:qn1}) indicate that the angular momentum $J$ becomes a
good quantum number for $J \ge 3$.  Indeed, for $J \ge 3$,
rotational lines belonging to different $J$ are energetically well
separated and the rotational splitting starts to dominate the
structure of the observed spectrum.

This can be understood as follows: A fast rotation of the
molecular  axis averages out  the direction of the electron spin
in the lab frame, because the electron spin is tightly coupled to
the molecular axis ($\Sigma = 1$). This leads to an effective
decrease of the Zeeman interaction. Similarly, a fast rotation
will prevent the nuclear spin $I$ from coupling to the molecular
axis (and therefore to the electron spin), as it cannot follow
fast enough in an adiabatic way. In fact, calculated expectation
values for $\Omega_I$  (see Table \ref{tbl:qn1}) are close to 0
(typically $|\langle\Omega_I\rangle| < 0.5$ for $J\geq3$)
indicating a strong averaging out. For slow rotations ($J \le 2$),
in contrast, the hyperfine and Zeeman interactions are of the same
order as the rotational splitting. This leads to a relatively
strong mixing between levels with $J = 1$ and $J = 2$.  Note that
for the $\clg$ manifold only levels with $J\geq 1$ exist since the
projection $\Omega = 1$.

The energetically lowest state has an expectation value
$\langle|\Omega_I|\rangle=2.6$,  which is close to its ``ideal"
value of 3 if the hyperfine interaction ($\propto\Omega_I \Sigma$)
was dominant over the rotational and Zeeman interactions. We can
now ask how large this hyperfine interaction is in the limit of
smallest rotation, e.g. $J = 1$.  We
can roughly estimate this by looking at Fig. 7. At $B 
 = 0$ the rotational
splitting of the $J=1$ and $J=4$ levels has to be
 18 $B_{\vib}$ = 7.2 GHz. The observed
splitting between the barycenter of the $J=4$ levels and the
lowest line at $\nu - \nu_o$ = 26.34 GHz is larger by about 2 GHz.
This indicates that the hyperfine structure for $J\cong1$ must be
spread out over a range of roughly 4\,GHz.

We now investigate the number of lines that can in  principle be
observed. For $J =4$ and $I = 1, 3$ we can form 9 states with $F
\ge 2$. For $ J = 3 $  we expect 8 states and for $J = 1, 2$
together we expect 10 states. These are the states which are
listed in Table II.  In contrast to the spectrum of $\cog$, we can
observe levels with even as well as odd $J$, due to a two-fold
degeneracy of each rotational level. One of the two levels has
positive parity while the other one (that is observed in the
experiment) has negative parity.

The fact that we observe levels with a rotation up to   $J = 4$
might at first be surprising, as we start from $J = 1$ $\Omega=0$
in the Feshbach state and the selection rule is  $\Delta J = 0,\pm
1$. However, this simply shows that $J$ in the excited state is
not a very good quantum number yet and the $J \approx 4$ levels
have a $J = 2$ contribution.

At zero magnetic field, the quantum number $F$ for the total
angular momentum is good. At this field we cannot observe levels
with $F > 3$, due to the selection rule $\Delta F = 0, \pm 1$. As
in Fig. 6, these levels are drawn with dashed lines. Again, at
larger magnetic fields levels with different $F$ mix, and as a
consequence more levels can be reached. The curves in Fig.
\ref{fig:omega1} also display  level crossings. Crossings are in
general avoided as long as the levels have the same $I$ quantum
number. For our specific Hamiltonian $H$, where $I$ is exactly
good quantum number, levels with different $I$ do not mix with
each other.

The overall structure of the observed spectroscopic lines is  well
described with our model, which essentially only has a single free
parameter, $b_F + {2 \over 3} c$.  However, we find significant
deviations of up to a few 100 MHz (e.g. note the data point next
to the small horizontal arrow in Fig. 7).  Such deviations clearly
lie outside our experimental uncertainty, especially since for
these cases relative positions of neighboring lines are determined
with a precision better than typically 30 MHz. On this level of
accuracy, in general, we do not have good agreement with the
theory.

In order to achieve better agreement, one must  include terms in
the theory that we have so far neglected. For example, the Zeeman
term $\propto \mathbf{L}\cdot \mathbf{B}$
 can
contribute in second order. It could also contribute to first
order if the projection $\Lambda $ is not completely vanishing due
to mixing-in of a $\Lambda = 1$ component  from nearby states,
e.g. via second order spin-orbit interaction. However, such
investigations would also require more detailed measurements and
must be left for future work.

\begin{table}[t]
\begin{center}
\renewcommand{\multirowsetup}{\centering} 
\begin{tabular}{|c||c|c|c|c|}
\hline
\multicolumn{5}{|c|}{$B 
=0$\,G}\\[1ex]
$\langle J\rangle$ & $\langle I\rangle$ & $|\langle \Omega_I\rangle|$ & $\langle F\rangle$ & $\nu-\nu_o$ (GHz)\\[1ex]
\hline
1.2 &   3.0 & 2.6 &  2.0 & 26.34\\
1.4 &   3.0 & 2.0 &  3.0 & 27.06\\
2.7 &   3.0 & 1.2 &  4.0 & 28.36\\
1.2 &   1.0 & 0.16&  2.0 & 28.53\\
2.0 &   3.0 & 1.0 &  2.0 & 29.00\\
2.0 &   3.0 & 0.41 &  3.0 & 29.68\\
1.9 &   1.0 & 0.27&  2.0 & 29.98\\
2.1 &   3.0 & 0.75&  5.0 & 30.12\\
2.3 &   1.0 & 0.031&  3.0 & 30.13\\
2.0 &   3.0 & 0.57&  4.0 & 30.51\\
\hline
3.0 &   3.0 & 0.75&  2.0 & 31.77\\
3.0 &   3.0 & 0.21&  3.0 & 32.02\\
3.0 &   1.0 & 0.11&  2.0 & 32.31\\
2.9 &   3.0 & 0.60&  4.0 & 32.44\\
3.0 &   1.0 & 0.025&  3.0 & 32.45\\
3.1 &   3.0 & 0.44&  6.0 & 32.56\\
3.2 &   1.0 & 0.069&  4.0 & 32.57\\
3.0 &   3.0 & 0.37&  5.0 & 32.64\\
\hline
3.9 &   3.0 & 0.13&  2.0 & 35.04\\
3.9 &   3.0 & 0.21&  3.0 & 35.41\\
4.0 &   1.0 & 0.014&  3.0 & 35.54\\
3.9 &   3.0 & 0.48&  4.0 & 35.78\\
4.0 &   1.0 & 0.13&  4.0 & 35.81\\
4.0 &   1.0 & 0.011&  5.0 & 35.99\\
3.9 &   3.0 & 0.54&  5.0 & 36.07\\
4.1 &   3.0 & 0.25&  7.0 & 36.19\\
4.0 &   3.0 & 0.30&  6.0 & 36.22\\
\hline
\end{tabular}
\caption{\label{tbl:qn1} Calculated $\clg$ $(\vib=13)$ excitation
frequencies  of the levels shown in Fig. \ref{fig:omega1} at 0\,G.
The offset from the excitation frequency $\nu$ is $\nu_o =
294600$\,GHz. The expectation values for quantum numbers $J$, $I$
and $F$ are also given. All levels have $S=1$, $\Sigma=1$, and
$M=2$. Levels with $I = 1$ are not observed.}
\end{center}
\end{table}

\section{Summary and Outlook}

\label{sec:con} In this article we carry out  high resolution
molecular spectroscopy starting with an ultracold  ensemble of
$^{87}$Rb$_2$ molecules. We are able to resolve the vibrational,
rotational, hyperfine and Zeeman structure of deeply bound states
($\vib < 15$) of the excited $\tripletex$ state. The accuracy of
the measured lines is about 60\,MHz, however, the relative
position of the lines with respect to each other is often
determined much better and its accuracy reaches 20 MHz. The
dominating feature of the observed spectra is the splitting of the
vibrational levels into a $\clg$ and a $\cog$ component which can
be understood as a strong effective spin-spin coupling of the
electrons. We obtain a good understanding of the level structure
to first order with a relatively simple effective model that only
takes into account the most important terms of the Hamiltonian. In
brief, the level structure for the $\cog$ line is mainly
determined by the vanishing spin component $\Sigma = 0$, which
leads to a very small hyperfine and Zeeman structure and a good
quantum number $J$. In contrast, for the $\clg$ line, the
hyperfine and Zeeman interactions are large for small rotations,
but then are averaged out at larger rotations such that the
rotational splitting according to $J(J+1)$ again determines the
spectrum. Despite the overall understanding of the level
structure, we still observe systematic deviations between
experiment and theory on the order of a few 100 MHz, which should
disappear in a more refined model. We find from our data that the
anisotropic part of the hyperfine interaction could essentially
not be determined. In order to do this, the experimental data of
the $\cog$ line will have to be measured with higher precision in
the future.

It would be interesting to see whether the determined fit
parameters for $\lambda$ (effective spin-spin interaction) and for
the hyperfine contact interaction can be deduced from ab-initio
calculations and the known atomic properties. Besides gaining a
better insight into the level structure of the so-far relatively
unexplored $\tripletex$ state, this work will be helpful for
future cold molecule experiments where the deeply bound molecules
need to be prepared with high efficiency in various well defined
quantum states of the $\triplet$ state. Levels in the $\tripletex$
state may then serve as an intermediate state-selective step in a
two-photon optical transfer scheme from Feshbach molecules
\cite{La08}.

\begin{acknowledgments}
T.T. C.S. and J.H.-D. thank Eberhard Tiemann for valuable
discussions on hyperfine interaction and angular momentum coupling
in molecules and for proofreading.  We are grateful to Rudi Grimm
for continuous generous support. We thank Gregor Thalhammer and
Klaus Winkler for early contributions to the measurements, and
Christiane Koch for helpful estimates of the Franck-Condon overlap
for the $\qmf - \qme$ transition. We thank Olivier Dulieu for
helping us identify the splitting of the vibrational levels into
$\cog$ and $\clg$ components. This work was supported by the
Austrian Science Fund (FWF) within SFB 15 (project part 17).
\end{acknowledgments}


\begin{thebibliography}{99}
\bibitem{expectval} In general, we define expectation values for the quantum
numbers of an operator as $\sum_i \lambda_i |\langle  e_i | \Psi
\rangle |^2$, where $\lambda_i$ are the eigenvalues of the
operator, $e_i$ are corresponding normalized eigenvectors, and $|
\Psi \rangle$ is the state.
\bibitem{lambdapar}
In order to obtain the splitting value of 47\,GHz we have taken
into account the full substructure of the $\cog$ and $1_g$ lines
and our effective Hamilton model. If we ignore the substructure
and simply use the observed barycenters of the $\cog$ and $1_g$
lines we obtain a splitting of about 42\,GHz.
\end{thebibliography}
\end{document}